\documentclass[12pt]{iopart} \usepackage{graphicx} \usepackage{amsfonts}
\usepackage{amssymb} \usepackage{amsthm} \usepackage{iopams}
\usepackage{setstack} \usepackage{dsfont} 
 
\begin{document} 
\title[Control of the central spin model]{Control of open quantum systems: \\Case study of the central spin model}.
\author{Christian Arenz$^1$, Giulia Gualdi$^2$ and Daniel
Burgarth$^1$} 
\address{$^1$ Department of Mathematics and Physics, Aberystwyth
University, Penglais Campus, SY23 2BZ Aberystwyth, Wales, United Kingdom}
\address{$^2$ Dipartimento di Fisica ed Astronomia, Universit\'a di Firenze, Via
Sansone 1, Sesto Fiorentino 50019 Italy; QSTAR, Largo Enrico Fermi 2, 50125
Firenze, Italy} 
\begin{abstract} 
We study the controllability of a central spin guided by a classical field and 
 interacting with a spin bath, showing that the central spin is fully controllable independently of
the number of bath spins. Additionally we find that for unequal  system-bath couplings even the bath becomes
controllable by acting on the central spin alone. We then analyze numerically
how the time to implement gates on the central spin scales with the number of bath spins 
and conjecture that for equal system-bath couplings it reaches a saturation value.
We provide evidence that sometimes noise can be effectively suppressed through control. 

\end{abstract} 

\section{Introduction}

The last decades have witnessed a spectacular
technological 
progress  to the extent that now the  
implementation of high-fidelity quantum technologies can be
thought of as a goal belonging
to the not-so-distant future.
However the loss of quantum coherence due to the 
unavoidable interaction of a quantum 
system with its surrounding environment \cite{Dec1}, i.e. decoherence, 
represents
the major obstacle on the way beyond proof-of-concept experiments
towards real-life functioning implementations. 

On the one hand
the quest for
a fundamental understanding of 
the sources and mechanisms  of decoherence attracts
substantial research effort,
while on the other the development
of strategies to minimize its detrimental effect
in view of practical applications
is also a major research focus.
Although  distinct these two research lines are deeply
intertwined since the deeper the understanding, 
the more effective
the strategies to fight decoherence can be. 
Within the context of quantum information
processing two prominent examples of strategies
against decoherence are
quantum error correction
and dynamical decoupling  \cite{ErrorC}.
However, the application of these schemes is subject
to some restrictions (e.g. logical overheads, noise threshold
or extremely short pulse timescales) 
and in general the implementation of a sequence of
quantum gates within these approaches remains hard. 
More flexible 
methods to counteract noise in such a way to allow
quantum computing and
in general survival of quantum coherence on useful timescales
are therefore highly desirable.
In this respect  quantum control theory offers a valuable way to go. 
The general idea behind quantum control is to
use the interaction
of a quantum system with a properly tailored classical
control field to
steer its dynamics towards the desired
outcome. 
In this perspective error correction and dynamical decoupling 
can be regarded as specific
instances of quantum control.

Unfortunately even without considering feedback schemes,
when it comes to open systems, quantum control  
is still a mathematically challenging subject.
Indeed although substantial
progress has been recently made \cite{OpenQSC1,OpenQSC2} to develop
an general framework, only very simple models have been solved so far. 
Our goal is to study a more rich system
by tackling the problem \emph{already
before the derivation of a reduced dynamics}. That is, we  
study the control system 
before the infamous "bath trace" is performed and
then conclude on the control properties of the open system. 
The problem with
this approach is that, for infinite baths, 
it requires the investigation of the control properties of
infinite dimensional systems which is almost equally challenging as that of open
systems. We have therefore decided to examine the spin-star model \cite{BoseSpinBath, DanielSpinBath, FischerSpinBath,SpinBathtwoOs}, where
a central spin interacts with a finite set of surrounding environmental spins. 
The spin star represents a finite but scalable
system, so that in principle the thermodynamic limit can be looked at,
and even for small environment sizes it is already experimentally relevant
because it can be used to describe the main sources of decoherence
in NV centers \cite{NVmodel, NVmodel2, NVQuantum1, NVQuantumC2} and quantum dots \cite{QubitSpinBath1, QubitSpinBath2, QuitSpinBath3}.
In addition the spin star model
has been subject to a series of studies
concerning its reduced dynamics, so there is hope to bring the two afore
mentioned research lines together on a practically relevant system.  
In the
present paper, we go in this direction 
by considering
a spin bath controlled through the central system, and completely characterizing
the theoretical control properties of both the bath and the central system.  


\section{The model} 
\label{sec:the model} 

\begin{figure}[htb] 
\centering
\includegraphics[width=200pt]{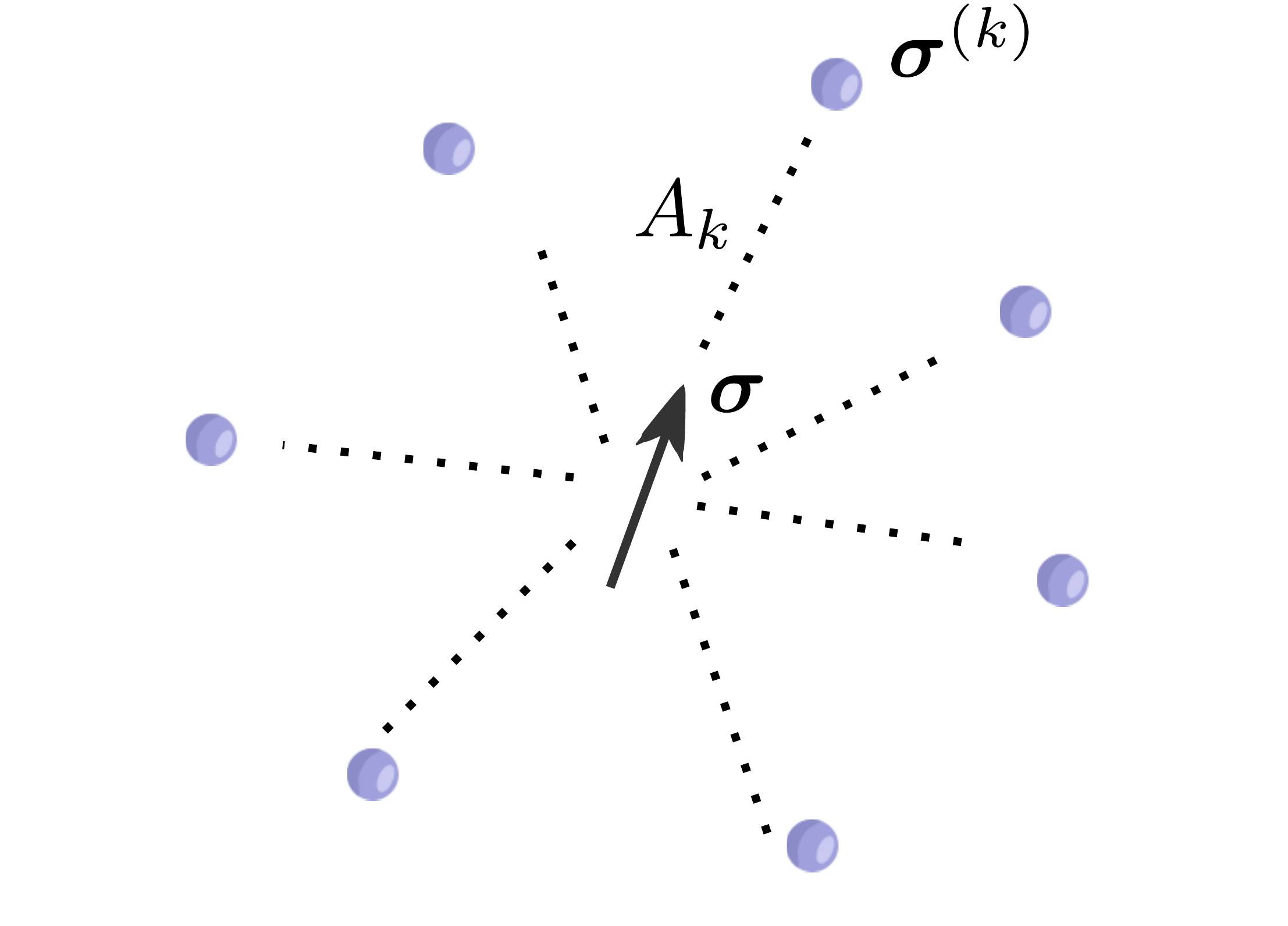}
\caption{\label{fig:SpinStar} The model described by Hamiltonian
\eref{eq:HamiltonianHeisenberg}: a central spin described by
$\boldsymbol{\sigma}$ interacts via an isotropic Heisenberg interaction with $N$
surrounding spins each described by $\boldsymbol{\sigma}^{(k)}$. The coupling
between the system and the $k$th bath spin is given by $A_{k}$. The central spin
interacts additionally with a classical control field as described by the
Hamiltonian \eref{eq:controllHam}.} 
\end{figure}

We consider a spin-star set up consisting of 
a central spin surrounded by $N$ spins as shown in Fig.\ref{fig:SpinStar}.
The spins surrounding the central spin will be hereafter referred to 
as the \textit{bath} spins keeping in mind that, strictly
speaking, they represent a true spin bath  only in the thermodynamic limit.
We assume that the central spin interacts with the bath spins via 
an isotropic Heisenberg interaction and that it is additionally subject to  
a constant magnetic field. The model is thus described
by the following Hamiltonian 
\begin{eqnarray}
H_{0}=\sigma_{y}+\sum\limits_{k=1}^{N}A_{k}
\boldsymbol{\sigma}\cdot\boldsymbol{\sigma}^{(k)},
\label{eq:HamiltonianHeisenberg} \end{eqnarray} 
where $A_{k}$ is the coupling
between the central and the $k$th bath spin, and
$\boldsymbol{\sigma}=(\sigma_{x},\sigma_{y},\sigma_{z})$ and
$\boldsymbol{\sigma}^{(k)}=(\sigma_{x}^{(k)},\sigma_{y}^{(k)},
\sigma_{z}^{(k)})^{T}$
are the Pauli matrices acting on the  central and the $k$th bath spin
respectively. Due to the isotropy of the Heisenberg interaction, the specific
choice of  $\sigma_y$ as the central spin Hamiltonian does not represent
a loss of generality.
Under the assumption
of equal system-bath couplings, i.e. $A_k = A$ for each $k$,
the dynamics of the central spin and the entanglement properties of 
similar models have been studied
analytically in \cite{BoseSpinBath, DanielSpinBath, FischerSpinBath} by means
of a non-Markovian master equation.  If all couplings are equal, in fact, 
the Hamiltonian
\eref{eq:HamiltonianHeisenberg} can be rewritten as a two-particle
Hamiltonian 
\begin{eqnarray}
H_{0}=i(\sigma_{-}-\sigma_{+})+2A(\sigma_{-}J_{+}+\sigma_{+}J_{-}+
\sigma_{z}J_{z}), \label{eq:HamiltonianCollectiveSpin} 
\end{eqnarray}
where $\sigma_{\pm}=(\sigma_{x}\pm i\sigma_{y})/2$ are the lowering and raising
operators acting on the central spin and the bath is regarded as 
a single effective particle with angular
momentum operator 
\begin{eqnarray}
\boldsymbol{J}=\frac{1}{2}\sum\limits_{k=1}^{N}\boldsymbol{\sigma}^{(k)},
\label{eq:angularmomentumop} 
\end{eqnarray}
and corresponding raising and lowering operators given by
$J_{\pm}=J_{x}\pm
iJ_{y}$. The Hamiltonian \eref{eq:HamiltonianCollectiveSpin}
conserves the square of the bath angular momentum, i.e.
$[\boldsymbol{J}^2,H_{0}]=0$. Hence, noting $[J_z,\boldsymbol{J}^2]=0$,
simultaneous eigenstates 
of $\boldsymbol{J}^2$ and $J_z$
represent a convenient basis for the bath. 
However, since the operators
$\boldsymbol{J}^{2}$ and $J_{z}$ alone do not form a complete set of
commuting observables, the subspaces defined by their eigenvalues, denoted by 
$j$ and $m$ respectively, are not in general one-dimensional. 
We therefore introduce an additional quantum number $\nu$
corresponding to the eigenvalues of certain permutation operators acting 
on the bath spins and commuting with $H_0$. 
The permutation operators do not need to be specified 
as the controllability analysis is
independent of them.  Due to the conservation of $j$ and $\nu$, the bath
Hilbert space can be written as a direct sum of the subspaces $\mathcal
H_{j,\nu}$ and the total Hilbert space can be written as $\mathcal
H=\mathcal H_{S}\otimes (\bigoplus_{j,\nu} \mathcal H_{j,\nu})$ where
$\mathcal H_{S}$ is the Hilbert space of the central spin.  This Hilbert
space structure, as detailed in the following section, lies at the heart
of the spin-star controllability properties in the equal coupling
scenario. 

Having defined the model Hamiltonian $H_0$, we now move on to
introduce controls. As discussed in the introduction, we assume that only 
the central spin can be accessed to and controlled. In order to obtain non-trivial
dynamics, the control field acting on the central spin  must not commute 
with $H_0$. A convenient choice is therefore represented by  a classical
magnetic  field $B(t)$ along the $z$ direction as described by the control
Hamiltonian \begin{eqnarray} H_{c}(t)=B(t)\sigma_{z}.
\label{eq:controllHam} \end{eqnarray}  
The full  Hamiltonian is thus \begin{eqnarray}
\label{eq:totalH} H(t)=H_{0}+H_{c}(t).  \end{eqnarray} 
Despite representing quite an extreme simplification, still the spin-star model 
described by Eq. \eref{eq:HamiltonianHeisenberg} already captures some relevant 
features of the  spin-bath decoherence processes occurring in solid-state systems
used for the implementation of quantum technologies  
such as nitrogen vacancy centers \cite{NVmodel,NVmodel2}
and quantum dots \cite{QubitSpinBath1,QubitSpinBath2, QuitSpinBath3}. Such a system therefore 
represents an interesting and challenging playground for an
investigation of controllability of open systems which can also be of practical
relevance. 

\section{Controllability considerations} 
\label{controllabilityspinstar} 
We now focus on the investigation of which unitary transformations
can be implemented on the spin star, in particular on the central spin,
using the control field $B(t)$.
The dynamics is governed by the Schr\"odinger equation for the time
evolution operator 
\begin{eqnarray}
\dot{U}(t)=-i(H_{0}+H_{c}(t))U(t),~~~~U(0)=\mathds{1},
\label{eq:schrödingereq}
\end{eqnarray} 
where the drift Hamiltonian $H_{0}$ and the control Hamiltonian
$H_{c}$ are those given in Eqs.  \eref{eq:HamiltonianHeisenberg} and
\eref{eq:controllHam}.  
All unitary operations which can be implemented on the
system constitute the \emph{reachable set} $\mathcal{R}$.  
More precisely
$\mathcal R$ is defined as the set containing all unitary transformations
$U$ which are solution of Eq.\eref{eq:schrödingereq} for some time $\tau>0$ and 
a certain control field
$B(t)$ with $t\in[0,\tau]$. The closure $\bar\mathcal R$ of the reachable set
consists of the unitaries which
can be achieved with arbitrary high precision.  
It is equal
\cite{BookDalessandro}  to the Lie
group $e^{\mathcal L}$, where $\mathcal L=\langle
iH_{0},iH_{c}\rangle_{[\cdot,\cdot]}$ is the \emph{dynamical Lie algebra} 
spanned by real linear combinations and nested commutators of $iH_{0}$ and
$iH_{c}$.  
The system is said to be fully  controllable if the Lie group is
equal to the unitary group or, in our case of traceless Hamiltonians, to the
special unitary group \cite{LieRankC1, LieRankC2}. To analyze the
controllability of the spin star we thus need to calculate the associated
dynamical Lie algebra.

Without bath spins, i.e. for $N=0$, the central spin is fully controllable
because $[i\sigma_{y},i\sigma_{z}]=2i\sigma_{x}$  and 
$\mathcal L=su(2)$.  When $N>0$
it is no longer obvious whether the central spin is fully controllable or not:
 on the one hand $H_0$ is necessary to achieve rotations around the $x$ axis, on 
the other the interaction with the bath spins introduces noise on the central 
spin. We will therefore study how the bath influences the controllability of
the central spin. The controllability of similar spin star models that consists of an anisotropic interaction of the central spin with the bath spins was studied in \cite{SimilarModelC1, SimilarModelC2}. Two classical fields were used to control the central spin and it was shown by using the graph criterion \cite{GraphCrit} that then the whole system becomes controllable. However this method is based on finding the eigenstates of the system and therefore it is not applicable for the Hamiltonian \eref{eq:HamiltonianHeisenberg}.
Moreover, in this case the controllability of the central spin is trivial, and can in principle be achieved arbitrarily quickly through strong control fields, which means that such models are not relevant in the
context of the present study.

In the following 
we will first consider the case when the central spin couples
with the same coupling strength to each bath spin 
and then the
case when the couplings are all different from each other.

\subsection{Equal couplings}
\label{chap_equalC} 
As discussed in section
\ref{sec:the model}, when the central spin couples to each bath spin with the
same strength, the bath spins behave like a collective spin described by the
angular momentum operator \eref{eq:angularmomentumop} whose square is conserved.
Since the control Hamiltonian \eref{eq:controllHam}  acts only on the central
spin, this symmetry is conserved also in presence of the control
field thus implying that the spin star is not fully controllable (see also
\cite{T.SchulteHSymmetries}). However, by performing repeated commutators of 
$iH_{0}$ and $iH_{c}$ and taking their real linear combinations, 
we can obtain the operators
$i\sigma_{\alpha},~iJ_{\alpha}~$ and $i\sigma_{\alpha}J_{\beta}$ with
$\alpha,\beta=x,y,z$ (see \ref{ref:ProofLieALEqualC} for details). 
This implies
that the full $su(2)$ algebra acting on the Hilbert space of the central spin is
contained in the dynamical Lie algebra regardless of the number of bath spins.
The central spin is thus fully controllable even in presence 
of decoherence or, in other words, the noise
induced on the central spin as a result of the interaction with the bath
can be effectively switched off.
More generally, 
the dynamical Lie algebra for equal
couplings contains all elements of the form (see \ref{ref:ProofLieALEqualC})
\begin{eqnarray}
\label{eq:liealequalC} i\sigma_{\alpha}(J_{+}^{l}J_{-}^{k}J_{z}^{s}+h.c.),~
~~\alpha= x,y,z~~l,k,s\in\mathbb N_{0}.
\end{eqnarray} 
Equation \eref{eq:liealequalC} implies full
controllability of the spin star within each subspace $\mathcal H_{S}\otimes (\bigoplus_{\nu} \mathcal H_{j,\nu})$
which can be achieved by properly combining the operators 
$J_{+}^{l}$ and $J_{-}^{k}$ in
such a way to act only on a given $j$-subspace.
Even without full controllability
it is still possible to perform many
interesting and practically relevant operations on the spin star
such as entangling the
central spin with the bath or using the bath as a data bus. Such protocols
were recently experimentally demonstrated in  \cite{NVcontroll_nuclear_spins}.
\\ 
The dimension of the dynamical Lie algebra can be obtained by determining the
size of the subspaces of fixed $\nu$ \cite{BoseSpinBath} as
$dim(\mathcal{L})=\sum_{j}((2(2j+1))^2-1)$.  
For a given $N$, $j$ can only take the values
$j=1/2,3/2,...,N/2$ when $N$ is odd and $j=0,1,...,N/2$ when $N$ is even,
we obtain 
\begin{eqnarray} 
\label{eq:dimensionLieal} 
dim(\mathcal{L})= 
\cases{ \frac{1}{6}(2+N)(9+4N(4+N)) ,  & \text{for}~N\text{ even},\\
\frac{1}{6}(1+N)(3+2N)(7+2N), & \text{for}~N\text{ odd},}
\end{eqnarray} 
which shows that the dimension of the dynamical Lie algebra
scales polynomially $\propto N^{3}$ with the size of the bath. 

\subsection{Different couplings} 

In the previous section we learned that in the
equal-coupling case the central spin is fully controllable but,
due to the symmetries of the system, the whole spin
star is not. The situation changes if all coupling constants $A_{k}$ 
are different from one
another. In this case the system has no more symmetries and  
the bath spins do not behave like a collective spin
anymore. Full controllability  of the central spin still holds for almost all
choices of the coupling constants and is independent of  both the size and the 
initial state of the bath, see (\ref{ref:proofVandermonde}). In addition each
single bath spin is fully controllable, see (\ref{sec:fullcontroll}), 
thus
allowing us to write 
\begin{eqnarray} 
\label{eq:star1}
\sigma_{\alpha}^{(k)}\in\mathcal L,~~~~\forall k=1,...N,~~\alpha=x,y,z.
\end{eqnarray} 
Hence, due to the Heisenberg interaction between the central and the bath spins,  
full controllability of the spin star $\mathcal
L=su\left(2^{N+1}\right)$ is achieved  \cite{localCgenreral}.
As a consequence, the dimension of the dynamical
Lie algebra scales exponentially  with the  bath size. 
By acting with a control field on the
central spin alone all degrees of freedom, even the unaccessible ones,
can be used for quantum information tasks.

\subsection{Implementing CPT maps} 

An interesting generalization of the above is to consider
the ability to implement
completely positive trace
preserving (CPT) maps on the central system. 
This is especially relevant in view
of the growing interest towards open quantum system simulators
\cite{opensym1,opensym2,opensym3,opensym4} and quantum reservoir engineering
\cite{QRE1,QRE2}.
We find that arbitrary CPT maps $\mathcal{D}(\rho_S)$ can be implemented: 
first, let us consider the unequal coupling case
with $N\ge2$. We initialize two spins of the bath in a pure state $\phi_{B}$ 
through consecutive unitary operations and
measurements on the central spin.  Using controls we then implement the unitary $U$ 
of the Stinespring representation
$\mathcal{D}(\rho_S)=\tr_{12}
\left\{U(\rho_S\otimes\phi_B)U^{\dagger}\right\}$ of $\mathcal{D}$, and thus $\mathcal{D}$.
Second, for  equal couplings even though the whole system is
not fully controllable it is still possible to implement every unitary operation
within the subspaces $\mathcal H_{S}\otimes (\bigoplus_{\nu} \mathcal H_{j,\nu})$. Provided they are large enough
($j>3/2$, implying $N>3$) and provided the bath can be initialized
appropriately, we can again implement a Stinespring dilation of $\mathcal{D}$. 

\subsection{Numerical calculation of the dynamical Lie algebra} 
\label{NumericsLieal}
\begin{figure} 
\begin{minipage}[hbt]{6cm} a) \\
\includegraphics[width=190pt]{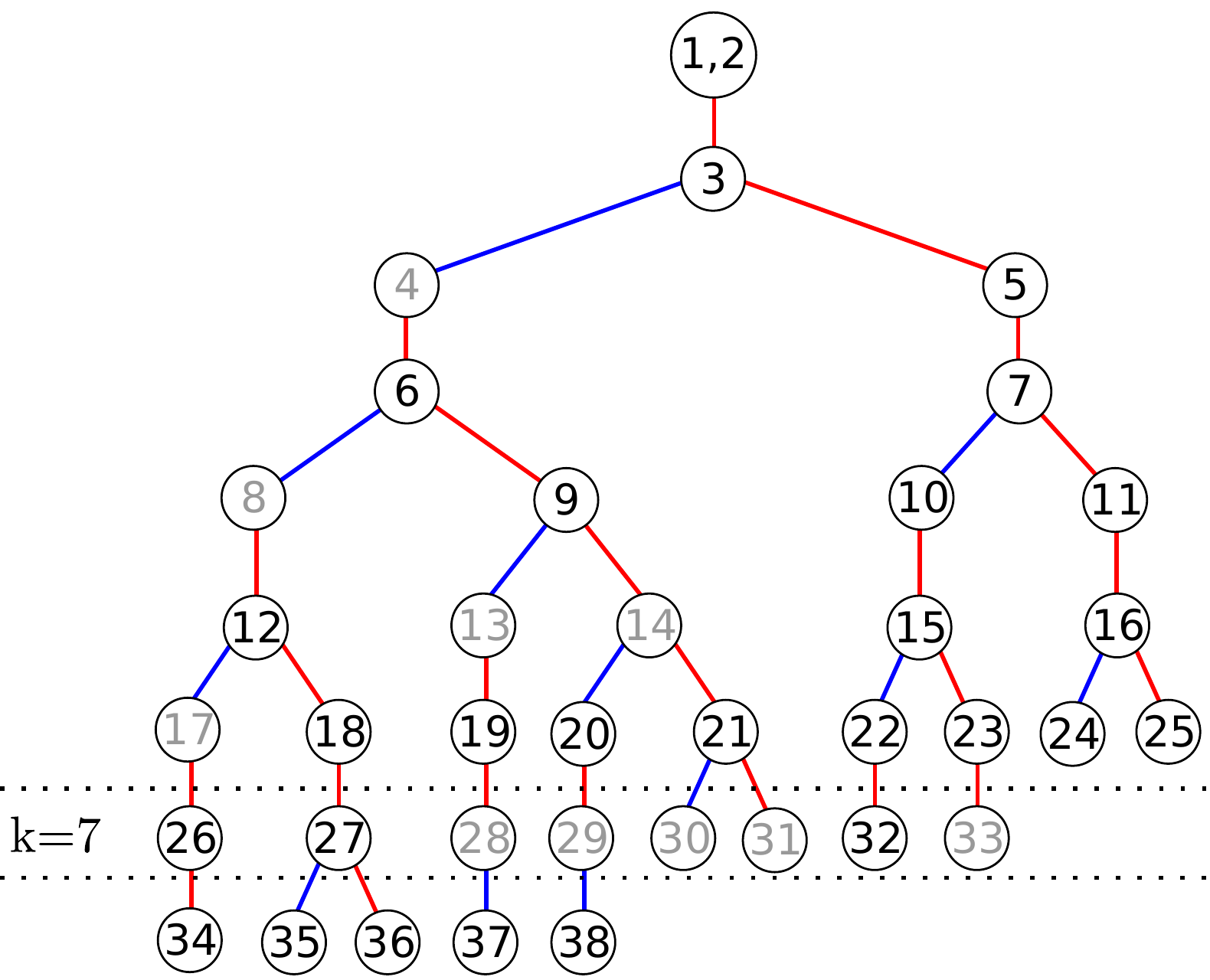} \end{minipage} \hfill
\begin{minipage}[hbt]{8.75cm} b)\\
\includegraphics[width=250pt]{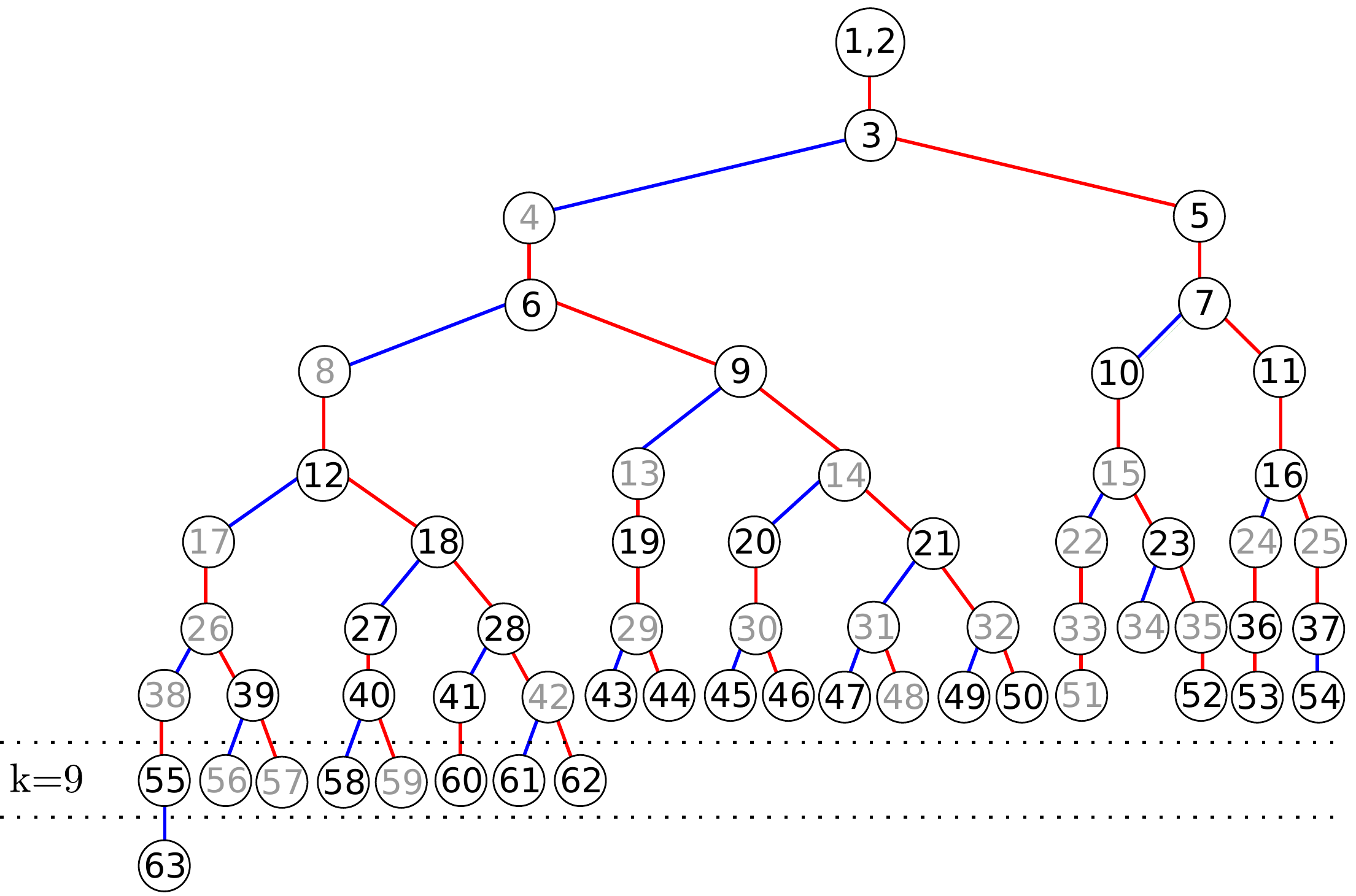} \end{minipage}
\caption{\label{fig:LieTree} Tree structure of the dynamical Lie algebra for
$N=2$ bath spins and either equal ($a)$) or different ($b)$) couplings. The numbers in the circles represent the elements of the
dynamical Lie algebra and the branches indicate whether the new linearly
independent term was obtained by commutation with $iH_{c}$ (blue) or
$iH_{0}$ (red). The index $k$ indicates the depth of the commutator.
Numbers in gray  denote the elements which, by real linear combinations, give
$i\sigma_x$ on the central spin such that  the central spin is fully
controllable.} 
\end{figure}

In this section we will examine more in detail the structure of the 
dynamical Lie algebra, $\mathcal{L}$, using a numerical algorithm similar to
those discussed in \cite{LieTreeAlg1} and  \cite{LieTreeAlg2}. 
In order to obtain
a complete operator basis for $\mathcal{L}$ it is enough to
repeatedly compute the commutators with $iH_{0}$ and $iH_{c}$, 
until the rank
of $\mathcal L$ does not increase any further  \cite{LieTreeAlg2}. Such
a procedure can be visualized as a tree, the so-called Lie tree. Indeed in Fig.
\ref{fig:LieTree} we show the Lie tree of  a spin star 
with $N=2$ bath spins for both equal, a),
and different, b), couplings. 
The numbers inside the circles label the elements
of $\mathcal{L}$ starting with $iH_{c}$ and $iH_{0}$ which
correspond to 1 and 2.  
The blue/red branches indicate that the new element was
obtained by commutation with $iH_{c}$/$iH_{0}$ respectively.
The number $k$ denotes the depth of the tree nodes
starting with $k=1$ for $[iH_{c}, iH_{0}]$. More generally, 
we define the depth of an element
of $\mathcal{L}$ as the maximal depth of nodes required to express it via linear combinations.  
Although the tree structure is not unique, because
it depends on the order according to which commutators are performed, using
the Jacobi identity it can be shown that the depth of an element is 
independent of the specific tree structure. 
To achieve full controllability of the central spin the crucial element to be
obtained is $i\sigma_x$. In order to determine its depth, we highlight nodes
corresponding to the basis elements that are needed to construct it in gray.

By comparing panels $a)$ and $b)$ of Fig. \ref{fig:LieTree}, where the depth of 
$\sigma_x$ is $k=7$ and $k=9$ respectively, we can conclude
that the value of the couplings, i.e. the presence of symmetries 
of the drift Hamiltonian, affects 
the depth at which full controllability of the
central spin is achieved. In both cases the tree structure is rather rich 
which is reflected by the complex proof of the central-spin full controllability 
presented in the appendix.

In the equal coupling case the depth of
$i\sigma_{x}$ is upper bounded by $24$ since it can be obtained in 
fashion independent of the bath size (see \ref{ref:ProofLieALEqualC}).  In contrast, for different couplings, the
depth of $i\sigma_x$
in the proof \ref{ref:proofVandermonde} indicates a linear scaling  with the bath size. 
However, this only represents an upper bound on the scaling because a different
proof might exist yielding a lower depth.
By considering a perturbation expansion of the time-evolution operator, it is
tempting to conjecture that the depth of an element of the dynamical Lie algebra
is related to the minimum time required to achieve its unitary companion.
Unfortunately we do not have enough numerical data to decide this conjecture and
leave it as an open problem for future studies.

\section{Influence of the bath on the minimum gate time}
\label{influencebathgatetime} 

So far we have discussed which unitary
transformations can be implemented \emph{in principle} on the spin star 
by a generic control field $B(t)$. By this we mean that no explicit statement
is made about the time required to achieve the desired unitary. In practice, 
we not only need to reach the desired unitary but we
need to do so in a reasonable time.
Therefore we now to turn to the
question of how the minimum time,  
$T^{*}(U_{G})$, required to implement
a target unitary transformation, $U_{G}$, on the central spin 
(hereafter minimum gate time) 
scales with the number of bath spins.  
To do this we need to identify the control pulse allowing to
implement $U_G$ in the shortest time
possible for different numbers of bath spins.  
To this end we need to resort to numerical gate optimization.

We  used the Grape  algorithm
\cite{Grape} as implemented in the open source optimal control package DYNAMO
\cite{webpageDYNAMO}. A detailed explanation of the algorithm and  the package
can be found in \cite{DYNAMOpaper}. The algorithm uses a gradient based method
that  maximizes the following gate fidelity 
\begin{eqnarray}
\label{eq:gatefid1} f_{1}(\tau)=\left|\frac{1}{2^{N+1}}
\text{tr}\{U_{G}^{\dagger}U(\tau)\}\right|^{2}, 
\end{eqnarray} 
given by the
modulus square of the normalized overlap, at a given time $\tau$,
between the target transformation, $U_G$, 
and the actual evolution, $U$.  
The fidelity
\eref{eq:gatefid1} involves choosing a  target unitary operation acting on the
whole spin star. However, since we are interested in implementing unitary
transformations on the central spin alone (and in general we cannot access the
bath degrees of freedom), such a choice is somewhat arbitrary and limiting.
In an open system set up a better and more motivated fidelity measure is
therefore given by \cite{GateFid1} 
\begin{eqnarray} 
\label{eq:fid2} 
f_2(\tau)= 1-\lambda\min_V||U_G\otimes V-U(\tau)||^{2}, 
\end{eqnarray} 
where $U_G$ is the
target unitary on the central system, $V$ a generic unitary on the bath, $U$
is the actual evolution at time $\tau$ of the full system and
$\lambda$ is a normalization constant. Due to the minimization
over all the unitaries acting on the bath, as opposed to the fidelity in Eq.
\eref{eq:gatefid1}, the fidelity defined in Eq. \eref{eq:fid2}, reaches its
maximum if the goal transformation has been implemented on the central system 
regardless of the bath evolution.  Using the Frobenius norm and choosing
$\lambda=1/(2*2^{N+1})$, the minimization can be
carried out explicitly yielding \cite{GateFid2}
\begin{eqnarray} 
\label{eq:gatefid2}
f_{2}(\tau)=\frac{1}{2^{N+1}}\text{tr}\{\sqrt{Q^{\dagger}Q}\},
\end{eqnarray} 
with
$Q=\text{tr}_{S}\{(U_G\otimes\mathds{1}_{\text{bath}})^{\dagger}U(\tau)\}$,
and $\text{tr}_{S}$ the partial trace over the central spin degrees of freedom.

After having included the gate fidelity $f_{2}$ into the DYNAMO package, we have
performed the
optimization of $f_{1}$ and $f_{2}$ by using the exact gradient
formula developed in \cite{ GateFid2,exactgradient}. The time
$\tau$, from now on called the driving time, has been divided into $M$ equidistant
time intervals $\Delta t=0,05$ chosen to be smaller than the inverse of the
highest eigenvalue of the Hamiltonian \eref{eq:HamiltonianHeisenberg} to
ensure a proper resolution of the dynamics.  
For a given bath size, in order
to estimate the minimum gate time $T^{*}$, we have  optimized both figures
of merit $f_{1}(\tau)$ and $f_{2}(\tau)$ for different values of $\tau$. We
additionally optimized over randomly chosen initial pulses meaning that at
each $\tau$ the maximum value of the fidelity over the different 
realizations is taken. An
additional optimization over many initial pulses is
performed to minimize the effect of local minima in the numerical routine.
Additionally, to ensure the 
existence of the thermodynamic limit
we have rescaled the coupling constants $A_{k}$ by
a factor $ \frac{A_{k}}{\sqrt{N}}$ \cite{FischerSpinBath}.

We emphasize that our numerical calculations can 
only provide upper bounds to the minimum
gate time, because the choice of initial control field can affect the
time at which the given fidelity reaches a predetermined threshold value. 

\subsection{Optimizing $f_{1}$} 

\begin{figure} \begin{minipage}[hbt]{8cm} a)\\
\includegraphics[width=220pt]{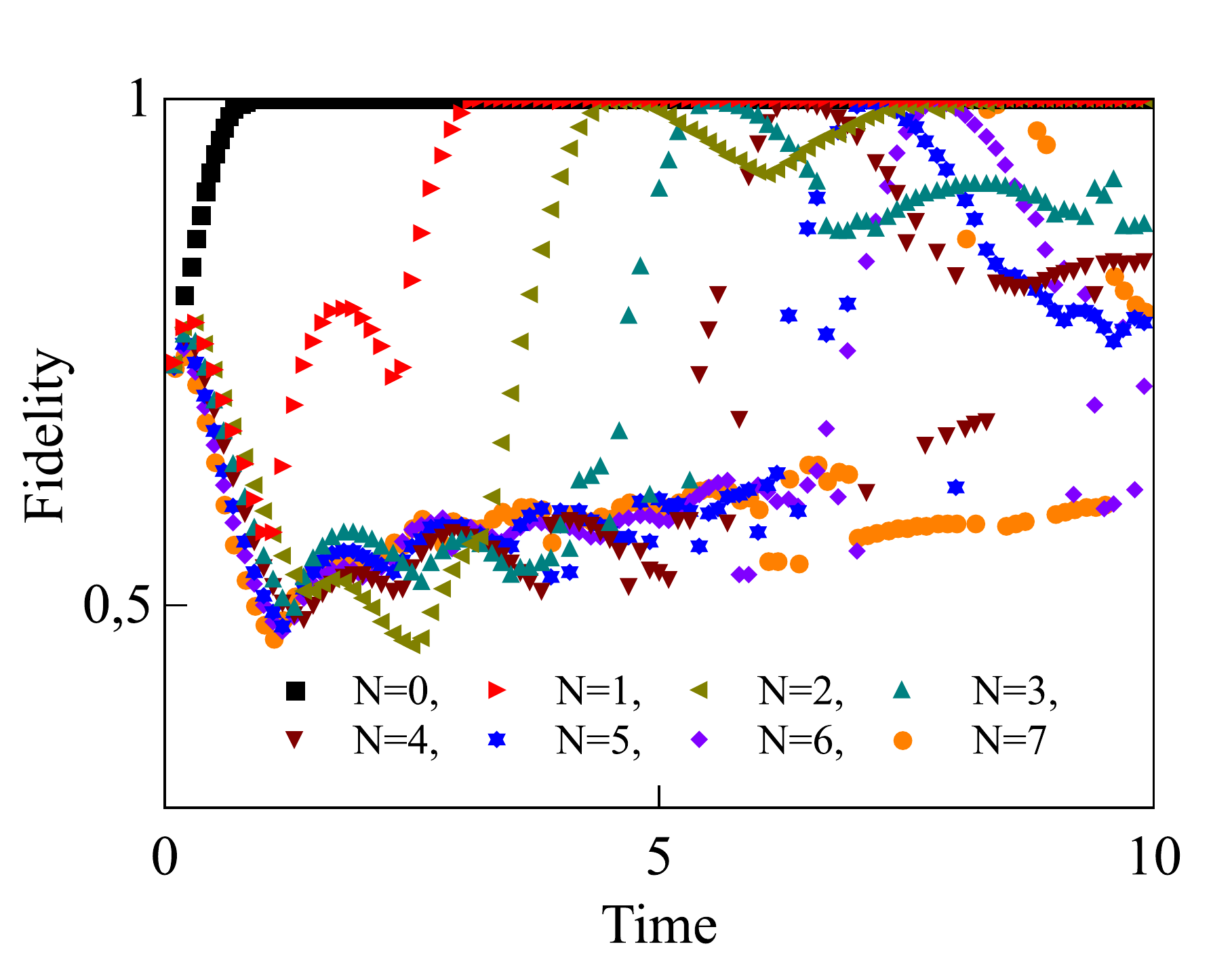} \end{minipage}
\hfill \begin{minipage}[hbt]{8cm} b)\\
\includegraphics[width=220pt]{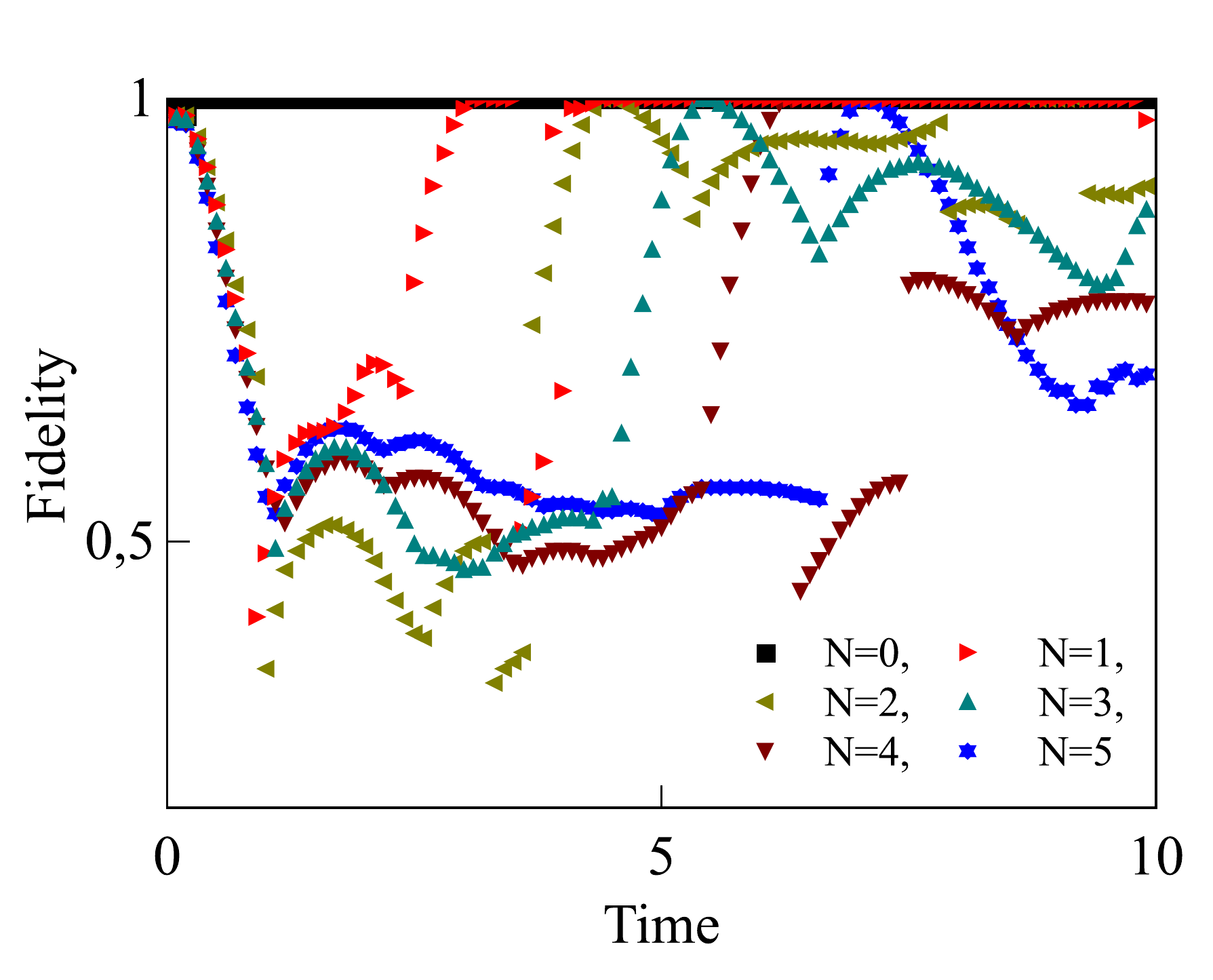}
\end{minipage} \caption{\label{fig:Fid_function_time_EC} Maximum value of the
fidelity $f_{1}(\tau)$  as a function of the driving time $\tau$ extracted
from 200 random initial pulses for equal couplings and target unitary $a)$
Hadamard on the central spin and identity on the bath; $b)$ $\pi/8$ on the
central spin and identity on the bath. Both plots have been obtained
for different numbers $N$ of bath spins as indicated on the figure.}
\end{figure}

\begin{figure}[h!] \centering
\includegraphics[width=220pt]{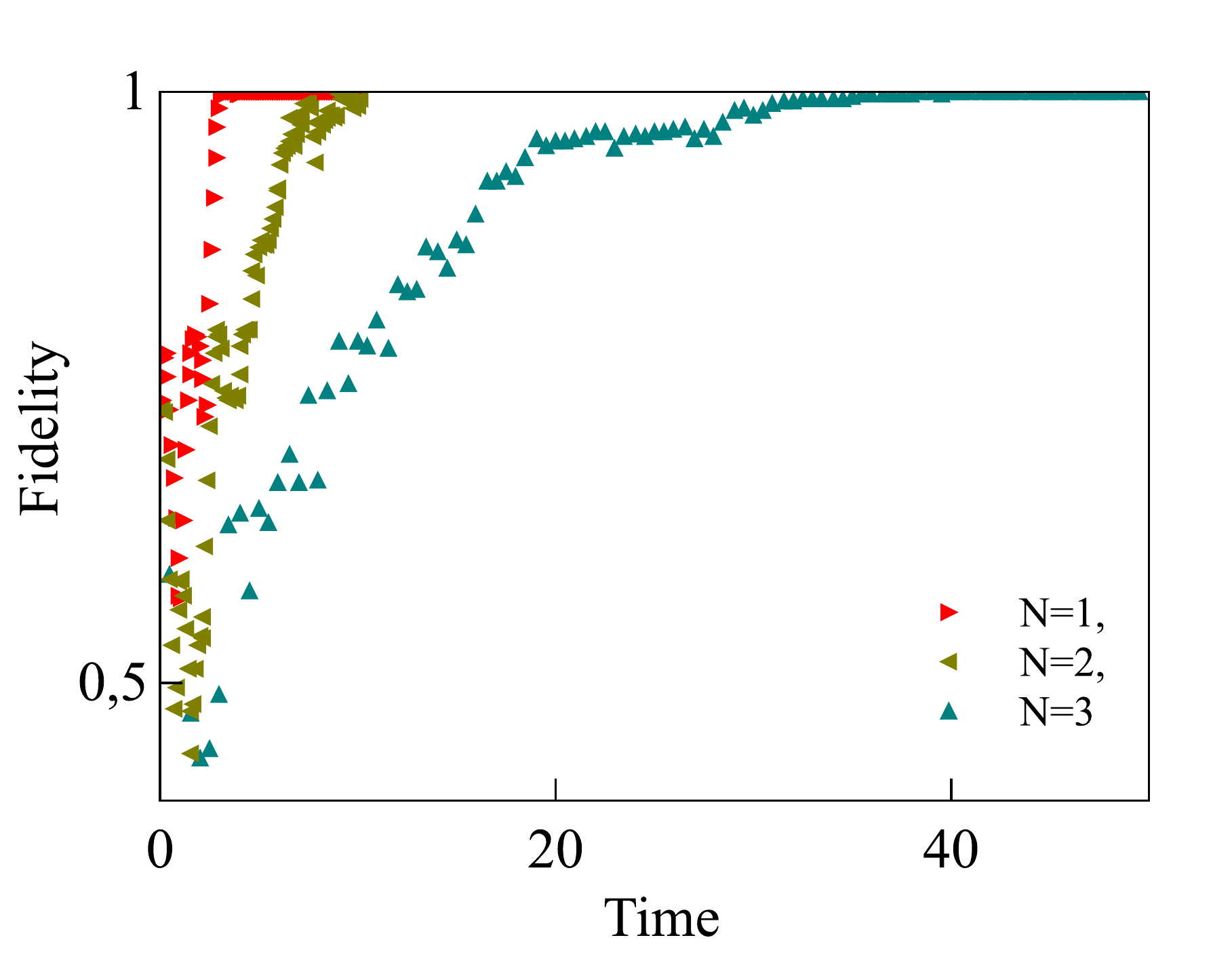}
\caption{\label{fig:Fid_function_time_DC} Maximum value of the fidelity
$f_{1}(\tau)$  as a function of the driving time $\tau$  extracted from 200
random initial pulses for different couplings and bath size $N=1,\cdots,3$.}
\end{figure}

We begin with the  optimization of the fidelity $f_{1}(\tau)$ defined in Eq. 
\eref{eq:gatefid1} and choose the identity as the target unitary on the bath. 
As a target transformation
$U_G$ on the central system we consider both the Hadamard gate and the $\pi/8$
gate since these one-qubit gates form a universal set \cite{NielseChuang}. 
We begin with  equal couplings, set $A=1$, and  investigate  the
minimum time required for the implementation of the Hadamard gate (Fig.
\ref{fig:Fid_function_time_EC} panel $a)$) and the $\pi/8$ gate (Fig.
\ref{fig:Fid_function_time_EC} panel $b)$) on the central spin.  In Fig.
\ref{fig:Fid_function_time_EC} we plot the maximum value of $f_{1}(\tau)$ as
a function of the driving time $\tau$ for different number of bath spins $N$
and maximized over 200 randomly chosen initial pulses.
Points that seem to break the continuity of the curves are statistical
fluctuations and have no physical relevance  as confirmed by optimizations over
a higher number of initial pulses. 
The computational effort required by the
optimizations is intensive which is the reason why, when optimizing $f_1$,
we restricted ourselves to a statistical sample of 200 random initial pulses for
each time and, when optimizing $f_2$, to 500. 
The black curve corresponds to
$N=0$ whereas the other curves to  increasing values of $N$: in panel $a)$
$N=1,\cdots,7$ and in panel $b)$ $N=1,\cdots,5$. We observe the following: \\ 
\textit{1) Short time behavior in the equal coupling case:} for
$\tau=0$ we have obtained $f_{1}(0)=0$ for the Hadamard gate and
$f_{1}(0)=(2+\sqrt{2})/4$ for the $\pi/8$ gate. The plots show the bath
detrimental effect on gate optimization on short time scales. Indeed after an
initial extremely short time window where all curves exhibit the same increasing
behaviour, reflecting the fact that correlations between the central system and
the bath have not been established yet, the  maximum value of the fidelities in
presence of the spin bath then drops compared to the $N=0$ case. Note that for
short times the $\pi/8$ gate can be reached with fidelities above $0.99$
independently of the number of bath spins. This reflects the fact that 
the $\pi/8$ gate is up to a global phase identical to a rotation around the 
$z$ axis which can
always be achieved at short times with a sufficiently large control-field 
amplitude.\\ 
\textit{2) Long time behavior in the equal coupling case:}
after a region of decreasing slope, all dissipative fidelities increase again
until, for all $N$,  a maximum value above $0.995$ is reached. The increasing
bath size results in a time shift of the  maximum value. 
The achieved maximum values are the same for both the Hadamard and the $\pi/8$
gate.  
\\ \textit{3) Different coupling case:} to study the effect of the
bath spins in more detail we will from now on focus only on the optimization of
the  Hadamard gate on the central spin.\\
Figure \ref{fig:Fid_function_time_DC}
shows the maximum value of the fidelity as a function of the driving time in the
different coupling case. The couplings are randomly chosen from
a uniform distribution between 1 and 2. The curves have been obtained for
$N=1,\cdots,3$ bath spins.  As before, a maximum value above $0.995$ is reached 
for all bath sizes but the driving time needed to reach it is much longer with
respect to the equal coupling case.  
\begin{figure}[h!] \centering
\includegraphics[width=220pt]{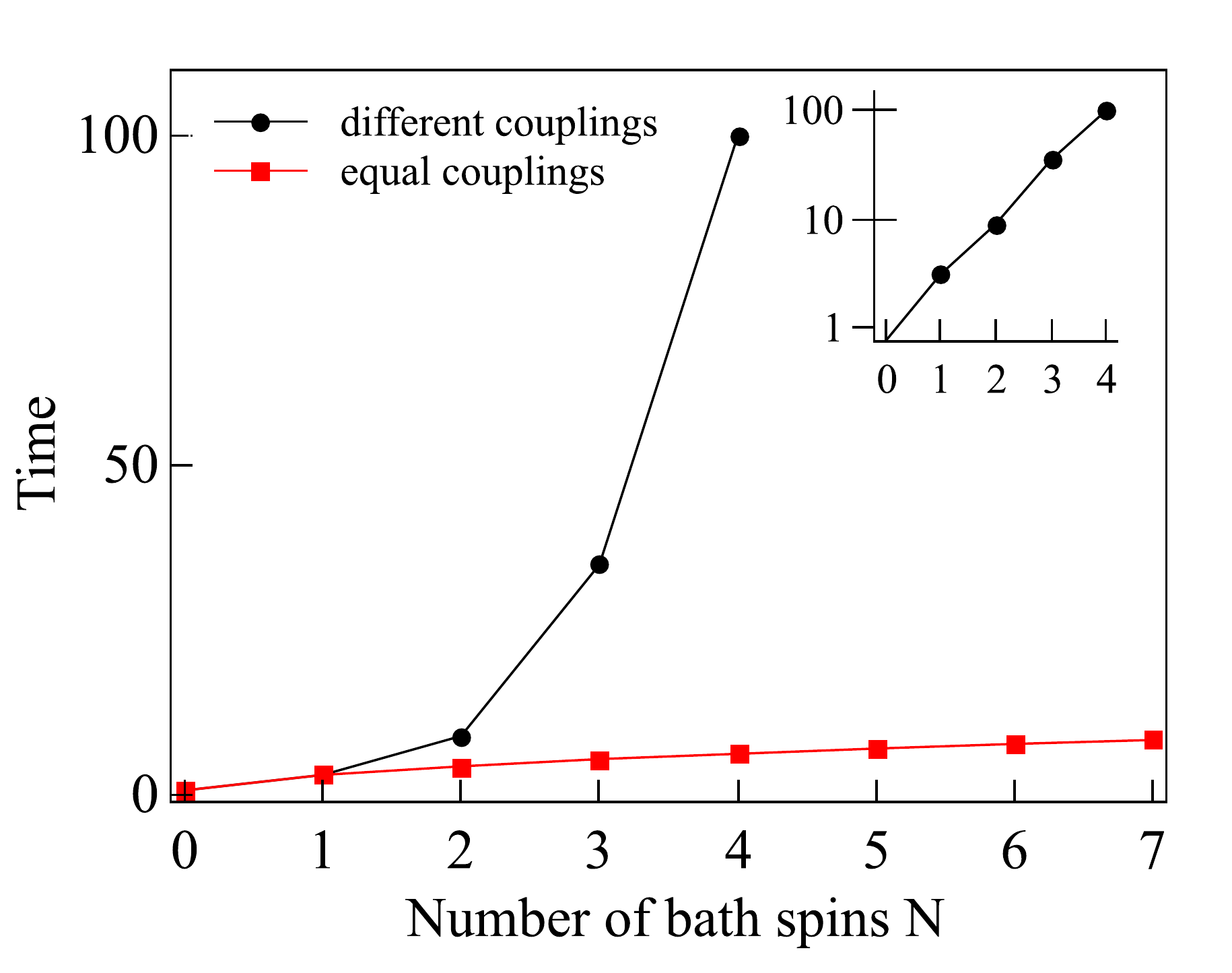}
\caption{\label{fig:Time_N} Minimum gate time $T^{*}$  needed to reach
a value of the fidelity of at least $f_{1}(\tau)=0.995$ as a function of the
number of bath spins $N$ for both different and equal couplings. The inset
shows the curve for different couplings on a logarithmic scale.} 
\end{figure}\\
\textit{4) Estimation of $T^{*}$:}  in Fig.
\ref{fig:Time_N} we plot the estimated minimum gate time $T^{*}$ 
against the number of bath spins for different and equal couplings.  Our
estimate has been obtained by setting 
a threshold value for the fidelity $f_{1}= 0.995$ and extracting the
corresponding $T^{*}$ from
the data plotted in Fig.  \ref{fig:Fid_function_time_EC} and
\ref{fig:Fid_function_time_DC}.  
The inset  shows the minimum gate time versus
the number of bath spins for different couplings on a logarithmic scale.  It
should be mentioned here that the point that belongs to $N=4$ for different
couplings was obtained by searching only in the expected time window for
a fidelity above the mentioned threshold. Furthermore, as already mentioned, our
results can only provide an upper bound on $T^{*}$.  Nevertheless Fig.
\ref{fig:Time_N} clearly suggests a significantly different scaling behaviour
of the minimum gate time in the two different coupling regimes. In the equal
coupling case, when the whole system is not fully controllable, the gate time
seems to depend weakly on the number of bath spins (red curve) in strong
contrast with the fully controllable case (black curve) where the dependence on
the bath size is at least polynomial (black curve).  Consistently with our
controllability analysis, the  scaling of the minimum gate time suggests that in
the equal coupling case the decoherence affecting the central spin can be
suppressed in reasonable time regardless of the size of the bath.
On the other hand, in the fully controllable case, for higher number of bath spins
(dramatically) longer gate times can be expected.  This seems consistent with
the intuition that if the dimension of the Lie Algebra grows exponentially with
$N$, then the implementation of a generic element of the corresponding Lie group
requires an exponentially increasing time.

\subsection{Optimizing $f_{2}$} Until now we have investigated the scaling of
the minimum gate time by optimizing $f_1$ and choosing the identity as a target
operation on the bath. We now want to see whether the optimization of $f_2$, Eq.
\eref{eq:gatefid2} exhibits significant deviations from
this behaviour.  
Unfortunately the optimization of $f_{2}$
with the GRAPE algorithm resulted extremely sensitive to local minima,
especially for increasing  number of bath spins, consistently also with the
results presented in \cite{GateFid2}.  In order to minimize this effect a much
higher number of random initial pulses was required. Due to computational
restrictions, we had therefore to limit 
our investigation to $N=2$ and 500 random initial
pulses. 
\begin{figure} \begin{minipage}[hbt]{8cm} a)\\
\includegraphics[width=220pt]{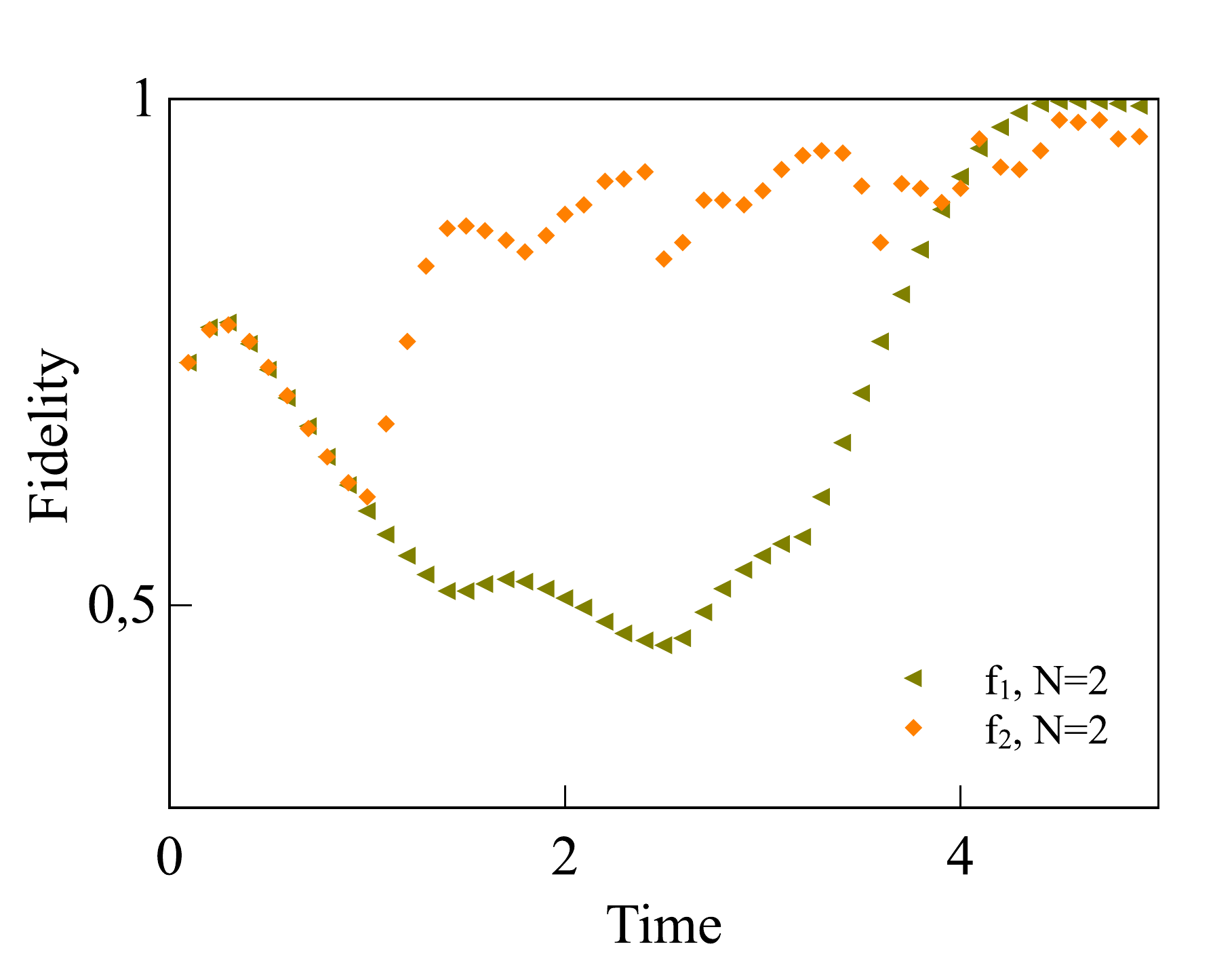}
\end{minipage} \hfill \begin{minipage}[hbt]{8cm} b)\\
\includegraphics[width=215pt]{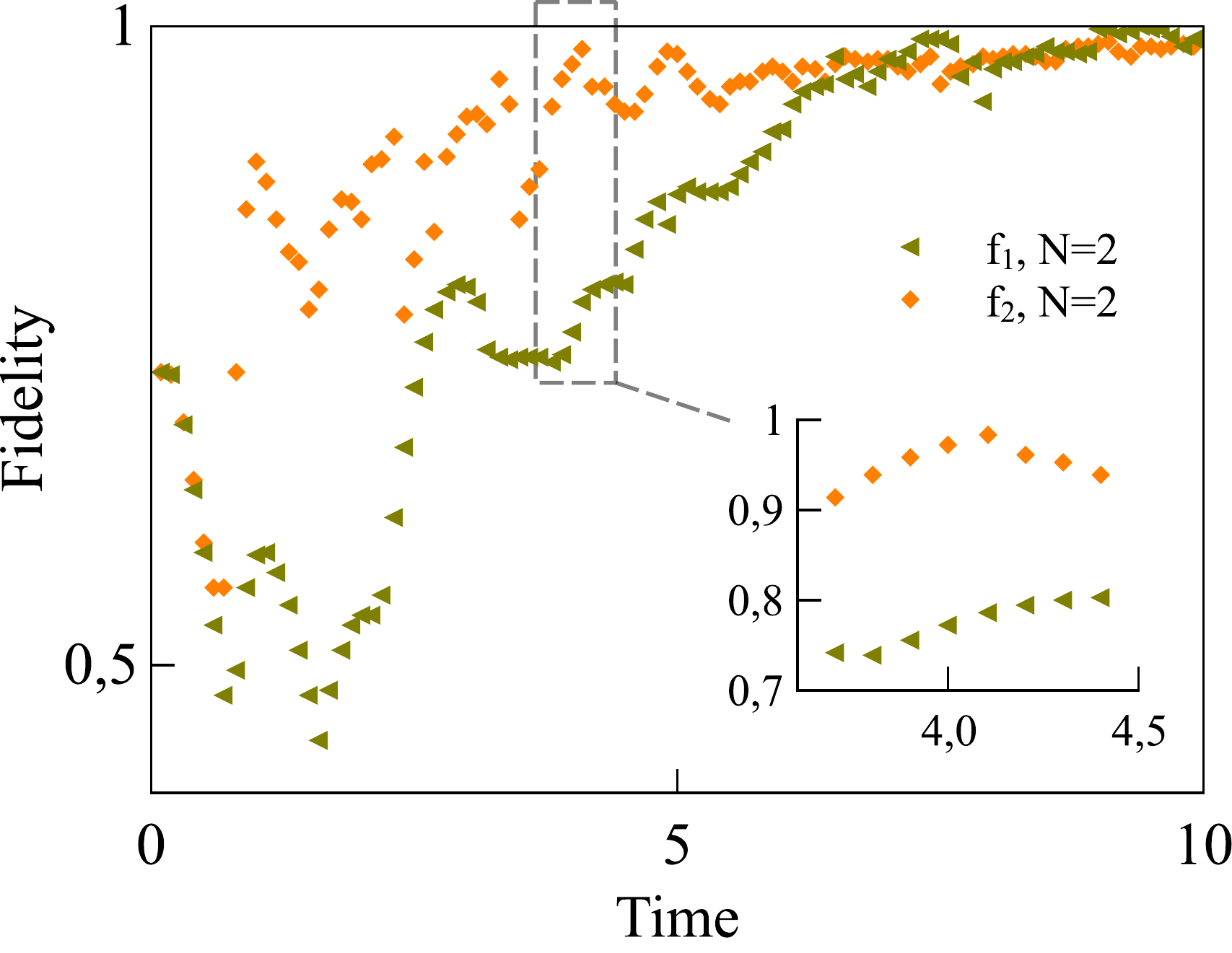}
\end{minipage} \caption{\label{fig:ig:GateFid_function_time_N2} Maximum values
of the fidelities $f_{1}(\tau)$ and $f_{2}(\tau)$ as a function of the driving
time $\tau$ extracted from 500 randomly chosen initial pulses for $N=2$: $a)$
equal couplings, $b)$ different couplings where the inset shows the time
window $\tau\in[3.7,4.4]$ and the maximum was extracted from $10^4$ random
initial pulses.} \end{figure}\\ 
In Fig.\ref{fig:ig:GateFid_function_time_N2} the
maximum value of the fidelities $f_{1}(\tau)$ (dark yellow curve) and
$f_{2}(\tau)$ (orange curve) is  plotted as a function of the driving time
$\tau$ for both equal (panel $a)$) and different couplings (panel $b)$).
Intuitively we would expect a shorter minimum gate time when the target
transformation is specified only on the central spin because in this case the
constraint on the bath evolution is weaker.  Each control pulse maximizing
$f_{1}(\tau)$ is a specific solution for $f_{2}(\tau)$ as well, hence the
fidelity $f_{2}(\tau)$ should at least attain the same maximum values as
$f_{1}(\tau)$.  However, from Fig.\ref{fig:ig:GateFid_function_time_N2}, we see
that values of the  maxima  reached by $f_2$ around $T^*$ are slightly below
those reached by $f_1$ thus witnessing an increased sensitivity of the
optimization to local minima within this time window.  
From Fig.\ref{fig:ig:GateFid_function_time_N2} we also note that up to a certain
time the curves relative to $f_{2}(\tau)$ and $f_{1}(\tau)$ are identical, thus
implying that within this time window there is no difference between setting the
target on the full system or on the central spin only. This behavior however
changes at increasing times since higher fidelities can be
achieved on shorter timescales if the target is only set on the central spin. 
Only at the end of
the time window the curves seems to become similar again. However, for equal
couplings, even though $f_2$ reaches higher values at shorter times, it never
crosses the threshold of $f_2=0.995$ before $T^*$ thus leading us to conclude
that in this case, for sufficiently high threshold values, 
the fidelity used does not significantly affect the estimate
of the upper bound on the minimum gate time. For different couplings, instead,
values that are close to the threshold can be reached at short times: the inset plot shows
a time window in which $f_2(\tau)$ reaches a  maximal value of  
$0.98$.

In conclusion, for equal couplings, the numerical results suggest that the minimum gate time
depends weakly on the size of the bath and perhaps reaches a saturation value. This is consistent to the theoretical prediction that the depth of an element of the dynamical Lie algebra is related to the minimum gate time to achieve its unitary companion (see section \ref{NumericsLieal}).
This behavior appears to be the same for both fidelities.  For different
couplings, instead, the numerical results hint at a  different behaviour of the
minimum gate time according to whether the target is defined on the whole system
or on the central spin only. In the latter case in fact not only the minimum
gate time seems to be shorter but also we can not even rule out the possibility
that it scales as in the equal coupling case.  More conclusive statements 
require much bigger computational resources and more sophisticated analytical
techniques as for example those suggested in \cite{timeopt1,timeopt2}. However these methods are not easily generalized to the high dimensional systems considered here.
      
\section{Conclusions} 
\label{conclusions} 

By analytical calculation of the dynamical Lie algebra, we showed that a 
central spin interacting with a surrounding spin bath is fully
controllable for almost all choices of the coupling constants and any bath size.
If the central spin couples to the bath with
unequal couplings, this property extends to the whole spin star, 
environmental spins included. We can 
therefore conclude that quite remarkably, by controlling the 
central system,
the bath can be i) effectively switched off; ii) arbitrarily engineered.
The possibility of controlling the environment via 
the central spin can be exploited to implement, on the central spin itself,
not only arbitrary unitaries but, more generally, arbitrary 
(completely positive trace preserving)
dynamical maps. This result can be of practical relevance 
both for quantum simulations of  open system and for quantum reservoir
engineering.

Alongside these purely analytical findings we also performed an extensive 
numerical investigation of control timescales and how these are
affected by both the bath size and the symmetries of the system.
In the maximally symmetric scenario, when all the bath spins can be regarded
as a single collective particle, our estimate for 
the minimum time required to perform a  gate under dissipative dynamics 
shows that it scales relatively slow, perhaps reaching
a saturation value, as a function of the bath size. On the contrary, 
in absence of symmetries, i.e. when each environmental spin interacts
differently with the central system, the scaling of the minimum gate time 
appears to be much faster (we conjecture exponentially faster).

Our results might have interesting applications in NV centers, 
which are essentially electron spins
in a finite nuclear spin bath. One recently demonstrated method to overcome the
short coherence time of the electron spin is to store its state in the nuclear
spins, which have longer decoherence times. Our control results then suggest that this might
not be the best strategy, and that instead one might apply a more complex shaped
pulse to the electron spin to keep it fresh for longer.

\ack The authors acknowledge fruitful discussions with Sophie
Schirmer, Shai Machnes, Thomas Schulte-Herbr\"uggen and Hubert de Guise. C.
Arenz acknowledges financial support through the Eleanor James Scholarship and
the Aberystwyth Doctoral Career Developement Scholarship.  G. Gualdi
acknowledges financial support from MIUR through the PRIN grant 2010LLKJBX, from
QYMRU, and through the IMAPS summer fellowship. QSTAR is the MPQ, LENS, IIT,
UniFi Joint Center for Quantum Science and Technology.  G. Gualdi and D.
Burgarth want to thank the KITP for support in part by the National Science
Foundation under Grant No. NSF PHY11-25915. We acknowledge the DYNAMO code \cite{webpageDYNAMO}.

\appendix  

\section{Lie algebra for equal couplings}
\label{ref:ProofLieALEqualC} 

First we want to show that
$i\sigma_{x}\in\mathcal L$ where we define  $A_{1}\equiv iH_{0}$ and
$A_{2}\equiv iH_{c}$.  \\ 
Building the double commutator $[A_{2},[A_{1},A_{2}]]$
we get up to a constant the element 
\begin{eqnarray}
A_{3}=i(\sigma_{y}+\sigma_{+}J_{-}+\sigma_{-}J_{+}), \end{eqnarray} which
leads with $A_{1}-A_{3}$ to \begin{eqnarray} A_{4}=i\sigma_{z}J_{z}.
\end{eqnarray} 
After calculating $[[A_{1},A_{2}],A_{3}]$ and using the
properties of $J_{-}$ and $,J_{+}$ we find up to a 
constant the element
\begin{eqnarray}
A_{5}=\sigma_{z}(J_{-}-J_{+})-2i\sigma_{z}(\vec{J}^{2}-
J_{z}^{2}-J_{z})+i\sigma_{-}\sigma_{+}J_{z}.
\end{eqnarray} 
The last two terms of $A_{5}$ commute with $A_{4}$
and therefore $[A_4,A_5]$ yields, up to a constant 
\begin{eqnarray}
A_{6}=iJ_{x}.  \end{eqnarray} 
By commuting $A_{6}$ with $[A_{1},A_{2}]$ we
find $i\sigma_{x}J_{z}$ and by commuting with $A_{2}$, $A_{4}$, $A_{6}$ we
obtain the following elements 
\begin{eqnarray}
&i\sigma_{z},~i\sigma_{x}J_{z},~i\sigma_{y}J_{z},~i\sigma_{z}J_{z},\nonumber\\
&iJ_{x},~i\sigma_{x}J_{y},~i\sigma_{y}J_{y},~i\sigma_{z}J_{y},
\label{eq:elements1} 
\end{eqnarray} 
which can be used to isolate
\begin{eqnarray} 
A_{7}=i(\sigma_{y}+\sigma_{x}J_{x}), 
\end{eqnarray} 
from $A_{1}$. 
By commuting $A_{7}$ with $i\sigma_{x}J_{z}$ we obtain $iJ_{y}$ and the commutator $[i\sigma_{x}J_{z},iJ_{y}]$
yields up to a constant $i\sigma_{x}J_{x}$ 
which can be used, together with with $A_{7}$, to reach $i\sigma_{y}$. 
We then also have $i\sigma_{x}$ by using $A_{2}$.
In fact we showed that 
\begin{eqnarray}
i\sigma_{\alpha},~iJ_{\beta},~i\sigma_{\alpha}J_{\beta}\in\mathcal
L,~~~~\forall \alpha,\beta=x,y,z.  \label{eq:liecentralbath}
\end{eqnarray} 
Due to the fact that the ladder operators $\sigma_{\pm}$
and $J_{\pm}$ define another representation it is easy to verify that
$i(\sigma_{\alpha}J_{\beta}+h.c.)\in\mathcal L$ holds also for
$\alpha,\beta=\pm, z$.\\ 
With the elements we found so far we can find
other elements by building their commutators and creating real linear
combinations. 
Next we show that
\begin{eqnarray}
i\sigma_{\alpha}(J_{+}^{l}J_{-}^{k}J_{z}^{s}+h.c.)\in \mathcal
L,~~~~\forall l,k,s\in\mathbb N,~~\alpha=x,y,z.  \label{eq:assumption}
\end{eqnarray} 
Essentially, this characterizes the dynamical Lie algebra up to normal ordering of operators.
We will proceed  by induction and define 
\begin{eqnarray}
A(K)=\text{span}\{i\sigma_{\alpha}(J_{+}^{l}J_{-}^{k}J_{z}^{s}+h.c.)~|~l+k+s\leq
K,~\alpha=x,y,z\}, \label{eq:A} 
\end{eqnarray} 
where hereafter Greek indices
describe some $x,y,z$ for the Pauli spin operators and some $\pm,z$
for the angular momentum operators. \\ 
The initial step is to
prove that $A(1) \subset\mathcal L$. This is trivial because we already
have proven with \eref{eq:liecentralbath} that
$\{i\sigma_{\alpha}(J_{\beta}+h.c.)\}$ is a subset of $\mathcal L$. We can
therefore go to the inductive step and 
show that if $A(K)\subset \mathcal L$ then
$A(K+1)\subset \mathcal L$.\\ 
Take any
$a=i\sigma_{\alpha}(J_{+}^{l}J_{-}^{k}J_{z}^{s}+h.c.)\in A(K+1)$ with
$l+k+s=K+1$ and calculate for $s>0$ the commutator 
\begin{eqnarray}
&[i\sigma_{\alpha}(J_{+}^{l}J_{-}^{k}J_{z}^{s-1}+h.c.),i\sigma_{\beta}J_{z}]
=\nonumber \\
&\sigma_{\beta}\sigma_{\alpha}(J_{z}J_{+}^{l}J_{-}^{k}
J_{z}^{s-1}+J_{z}^{s}J_{+}^{k}J_{-}^{l})-
\sigma_{\alpha}\sigma_{\beta}(J_{+}^{l}
J_{-}^{k}J_{z}^{s}+J_{z}^{s-1}J_{+}^{k}J_{-}^{l}J_{z}),
\label{eq:commutat} \end{eqnarray} 
keeping in mind that
if $i\sigma_{\alpha}(J_{+}^{l}J_{-}^{k}J_{z}^{s-1}+h.c.)\in A(K)$ then
the above commutator is by construction an element of $\mathcal L$. 
Due to the anticommutation rules of the Pauli spin operators, we can always
choose a $\sigma_{\beta}$ so to obtain from
Eq. \eref{eq:commutat} up to a constant the following
\begin{eqnarray}
[i\sigma_{\alpha}(J_{+}^{l}J_{-}^{k}J_{z}^{s-1}+h.c.),i\sigma_{\beta}J_{z}]
=a+O, \end{eqnarray} with $O\in A(K)$. 
The cases $l>0$ and $k>0$
can be treated analogously and therefore we showed that $a\in
\mathcal L,~\forall l,k,s\in\mathbb N$. $\qed$

\section{Controllability proofs} 

\subsection{Controllability of the central spin} 
\label{ref:proofVandermonde} 

In this section we will prove controllability
of the central spin by using the determinant of a Vandermonde matrix along the
lines of \cite{Vandemonde}.  We want to prove that $su(2)\subset\mathcal L$,
$\forall N\in\mathbb N$  for almost all values of the couplings constants
$A_{k}$. By $su(2)$ we denote the special unitary algebra acting on the central
spin.  To be as general as possible we rewrite the system Hamiltonian
\eref{eq:HamiltonianHeisenberg} as
\begin{eqnarray} 
\label{eq:macroham}
H_{0}=\sigma_{y}+\sum\limits_{n=1}^{\tilde N}h_{n}(\sigma_{x}J_{x}^{(n)}+
\sigma_{y}J_{y}^{(n)}+\sigma_{z}J_{z}^{(n)}), 
\end{eqnarray} 
where each set $n$ of bath
spins with identical $A_{k}$'s are combined as collective particles,
coupled to the central system with strength $h_{n}$ and with corresponding
angular momentum operators $J_{\alpha}^{(n)}$ with $\alpha=x,y,z$.
We assume that $|h_{n}|\neq |h_{m}|$ and $|h_{n}-h_{m}|\neq |h_{i}-h_{j}|$  with
$(n,m)\neq (i,j)\neq (j,i)$. In general these assumptions are only instrumental 
to the analytical proof and have neither physical
meaning nor are necessary in practice 
as witnessed by numerical calculations of the dimension
of the dynamical Lie algebra. An exception occurs for the full controllability 
of the whole spin star. In this instance, which will be discussed later 
in \ref{sec:fullcontroll}, both the
analytical proof and the numerical calculations show that the
assumption $|h_{n}|\neq |h_{m}|$ is necessary. \\ 
In order to prove full
controllability of the central spin, we need to prove that the operator
$i\sigma_x$ acting on the central spin belongs to the dynamical Lie
algebra $\mathcal L$.  To this end we begin by commuting $iH_{0}$ with the
control Hamiltonian \eref{eq:controllHam} and get by real linear combinations
the elements
\begin{eqnarray} 
B_{1}&=i(\sigma_{y}+\sum\limits_{n=1}^{\tilde
N}h_{n}(\sigma_{x}J_{x}^{(n)}+ \sigma_{y}J_{y}^{(n)})),\\
B_{2}&=i\sigma_{z}\sum\limits_{n=1}^{\tilde N}h_{n}J_{z}^{(n)},\\
B_{3}&=i(\sigma_{x}+\sum\limits_{n=1}^{\tilde N}h_{n}(\sigma_{x}J_{y}^{(n)}-
\sigma_{y}J_{x}^{(n)})). 
\end{eqnarray} 
We can now observe that proving
$i\sigma_x\in\mathcal L$ amounts to prove that $iJ_x^{(i)}\in\mathcal L$. Indeed
$i\sigma_x$ is obtained by performing commutators of $iJ_{x}^{(i)}$ and $B_{}1$, $B_{2}$, $B_{3}$ and real linear combinations of the resulting elements. 
The double commutator $[[B_{1},B_{3}],B_{2}]$ yields up 
to a constant the
element 
\begin{eqnarray} 
B_{4}&=i(\sum\limits_{n=1}^{\tilde
N}h_{n}^{2}J_{x}^{(n)}\nonumber\\ &+\sum\limits_{n>m=1}^{\tilde
N}(h_{n}-h_{m})h_{n}h_{m} (J_{x}^{(n)}J_{y}^{(m)}-J_{x}^{(m)}J_{y}^{(n)})).
\end{eqnarray} 
At this point the key observation is that up to a
constant
\begin{eqnarray} 
\label{eq:ob1} 
&\sum\limits_{n>m=1}^{\tilde
N}c_{n,m}[[(J_{x}^{(n)}J_{y}^{(m)}-J_{x}^{(m)}J_{y}^{(n)}),
B_{2}],B_{2}]\nonumber \\ =&\sum\limits_{n>m=1}^{\tilde
N}(h_{n}-h_{m})^{2}c_{n,m}(J_{x}^{(n)} J_{y}^{(m)}-J_{x}^{(m)}J_{y}^{(n)}),
\end{eqnarray} 
and 
\begin{eqnarray} \label{eq:ob2} \sum\limits_{n=1}^{\tilde
N}d_{n}[[J_{x}^{(n)},B_{2}],B_{2}]=\sum \limits_{n=1}^{\tilde
N}h_{n}^{2}d_{n}J_{x}^{(n)}, 
\end{eqnarray} 
with $c_{n,m}$ and $d_{n}$ some
coefficients.  Using the operator $B_{4}$ and Eqs.  \eref{eq:ob1} and
\eref{eq:ob2}, we can create operators of the form  
\begin{eqnarray}
\label{eq:operator1} 
B^{(s)}&=i(\sum\limits_{n=1}^{\tilde
N}h_{n}^{2(s+1)}J_{x}^{(n)} \nonumber\\ &+\sum\limits_{n>m=1}^{\tilde
N}(h_{n}-h_{m})^{2s+1}h_{n}h_{m}(J_{x}^{(n)}J_{y}^{(m)}-
J_{x}^{(m)}J_{y}^{(n)})), 
\end{eqnarray} with
$B^{(0)}=B_{4}$,~$[[B^{(s)},B_{2}],B_{2}]=B^{(s+1)}$ and $s=0,...,\tilde N-1$.
\\ 
We now need to show that the operators of the kind
$X^{(s)}\equiv\sum\limits_{n=1}^{\tilde N}h_{n}^{2(s+1)}J_{x}^{(n)}$ contained
in each $B^{(s)}$, Eq. \eref{eq:operator1}, are all linearly independent. In
fact, if
all $X^{(s)}$ are linearly independent then the determinant of the matrix
corresponding to  the linear set of equations 
\begin{eqnarray} 
\label{matrix}
\left( \begin{array}{cccc} h_{1}^{2} & h_{2}^{2} & \cdots  & h_{\tilde
N}^{2}\\ h_{1}^{4} & h_{2}^{4} & \cdots & h_{\tilde
N}^{4}\\ \vdots & \vdots &  \ddots & \vdots\\
h_{1}^{2\tilde N} & h_{2}^{2\tilde N} & \cdots
& h_{\tilde N}^{2\tilde N} \end{array}
\right) \left( \begin{array}{c} J_{x}^{(1)}\\ J_{x}^{(2)}\\ \vdots\\
J_{x}^{(\tilde N)} \end{array} \right)= \left( \begin{array}{c} X^{(0)}\\
X^{(1)}\\ \vdots\\ X^{(\tilde N-1)} \end{array} \right) 
\end{eqnarray}
is non-vanishing.  We now define $\tilde {h}_{n}= h_{n}^{2}$,
divide the columns of the matrix \eref{matrix} by $\tilde h_{n}$ and
then transpose. In this way we obtain a Vandermonde
matrix whose  determinant $\prod\limits_{1\leq i < j\leq \tilde
N}(\tilde{h}_{j}-\tilde{h}_{i})$ is non-vanishing if $|h_{j}|\neq
|h_{i}|$, $\forall i\neq j$  as assumed in the beginning. \\ 
By real
linear combination of the operators $B^{(s)}$ the operator
\begin{eqnarray} 
\label{eq:operator2} 
\tilde B_{i}=i(J_{x}^{(i)}+\sum\limits_{n>m=1}^{\tilde
N}\xi_{n,m}^{(i)}(J_{x}^{(n)}J_{y}^{(m)}- J_{x}^{(m)}J_{y}^{(n)})),
\end{eqnarray} 
can be selected. If all $\xi_{n,m}^{(i)}$ in Eq.
\eref{eq:operator2} are zero then we immediately obtain the
operator $iJ_{x}^{(i)}$ as an element of $\mathcal L$.  
If this is not
the case, using Eqs. \eref{eq:ob1} and \eref{eq:ob2} we can again
construct $s$ operators of the form 
\begin{eqnarray}
\chi^{(s)}&=i(h_{i}^{2s} J_{x}^{(i)} \nonumber\\
&+\sum\limits_{n>m=1}^{\tilde N}
(h_{n}-h_{m})^{2s}\xi_{n,m}^{(i)}
(J_{x}^{(n)}J_{y}^{(m)}-J_{x}^{(m)}J_{y}^{(n)}))  
\end{eqnarray} with
$s=1,...,(\tilde N^{2}-\tilde N)/2$ assuming that all coefficients
$\xi_{n,m}^{(i)}$ are different from zero.  As before we can associate
them to a Vandermonde matrix with non-vanishing determinant provided
that $|h_{n}-h_{m}|\neq |h_{i}-h_{j}|,~~\forall (n,m)\neq (i,j)\neq
(j,i)$. By real linear combinations of the $\chi^{(s)}$'s  we can then
select the operator 
\begin{eqnarray} 
\tilde \chi_{n,m}=i(\lambda
J_{x}^{(i)}+\omega_{n,m}(J_{x}^{(n)}J_{y}^{(m)}-J_{x}^{(m)}J_{y}^{(n)})).
\end{eqnarray} 
If the coefficient $\lambda$ is zero we can obtain
$iJ_x^{(i)}$ by real linear combinations of
$\omega_{n,m}(J_{x}^{(n)}J_{y}^{(m)}-J_{x}^{(m)}J_{y}^{(n)})$ and the
$B_{i}$'s  \eref{eq:operator2}.  Instead, if $\lambda\neq 0$, using Eqs.
\eref{eq:ob1} and \eref{eq:ob2}, we can obtain from $\tilde\chi_{n,m}$
a second linearly independent operator with the same structure and then,
by real linear combination of the two operators, the operator
$iJ_{x}^{(i)}$. Since $iJ_x^{(i)}\in\mathcal L$ we have
$i\sigma_x\otimes\mathds{1}_{bath} \in\mathcal L$ and hence the central
spin is fully controllable.

\subsection{Full controllability} \label{sec:fullcontroll} 

By commuting $iJ_{x}^{(i)}$ with $B_{1}$ and $B_{2}$ and using the full controllability of the central spins we obtain by real linear
combinations $iJ_{y}^{(i)}\in\mathcal L$ and hence
$iJ_{z}^{(i)}\in\mathcal L$. This implies that each collective particle
contained in Eq. \eref{eq:macroham} is fully controllable.  If all system-bath
coupling constants are different from each other this implies full
controllability of each bath spin and due to the Heisenberg interaction with the
central spin the Lie algebra is given by $su(2^{N+1})$ \cite{localCgenreral}
meaning that the whole system is fully controllable. We emphasize that
controllability of the whole spin star can only be achieved if all coupling
constants are different from each other, because in this case the existence of
symmetric manifolds is prevented. The numerical calculation of the dimension of
the dynamical Lie algebra shows that even the absolute value of the coupling
constants has to be different from each other.

\end{document}